%% file: LGdwarfsHubble_paper.tex
\documentclass[fleqn,usenatbib]{mnras}

\usepackage{mathptmx}
\input{embed_video}
\newif\ifembedvideo
\embedvideotrue  %\embedvideofalse to toggle
\usepackage[T1]{fontenc}

\DeclareRobustCommand{\VAN}[3]{#2}
\let\VANthebibliography\thebibliography
\def\thebibliography{\DeclareRobustCommand{\VAN}[3]{##3}\VANthebibliography}

\usepackage{graphicx}	% Including figure files
\usepackage{amsmath}	% Advanced maths commands

\usepackage{url,times,graphicx,hyperref,amsmath,amsfonts,amssymb,aas_macros,color,epsfig,epstopdf}
\usepackage{booktabs}
\usepackage[utf8]{inputenc}
\usepackage{ae,aecompl}
\usepackage{subcaption}

\title[The anisotropic distribution of dwarfs in the Local Group and beyond]{Anisotropies in the spatial distribution and kinematics of dwarf galaxies in the Local Group and beyond}

\author[Santos-Santos, Navarro \& McConnachie]{
Isabel M.E.  Santos-Santos,$^{1}$\thanks{E-mail: isabel.santos@durham.ac.uk}
Julio F.  Navarro,$^{2}$
Alan McConnachie$^{3,2}$
\\
$^{1}$Institute for Computational Cosmology, Department of Physics, Durham University, South Road, Durham, DH1 3LE, UK\\
$^{2}$Department of Physics and Astronomy, University of Victoria, Victoria, BC V8P 5C2, Canada\\
$^{3}$NRC Herzberg Astronomy and Astrophysics, 5071 West Saanich Road, Victoria, BC, V9E 2E7, Canada
}

\date{Accepted XXX. Received YYY; in original form ZZZ}

\pubyear{2023}

\begin{document}
\label{firstpage}
\pagerange{\pageref{firstpage}--\pageref{lastpage}}
\maketitle

\begin{abstract}
  The Local Group (LG) of galaxies is dominated by the Andromeda (M31) and Milky Way (MW) pair, a configuration which suggests that the mass distribution in the LG and its surroundings should be highly anisotropic. We use the APOSTLE suite of cosmological hydrodynamical simulations to examine how this anisotropy manifests on the spatial distribution and kinematics of dwarf galaxies out to a distance of $\sim 3$ Mpc from the MW. The simulations indicate a clear preference for dwarfs to be located close to the axis defined by the MW-M31 direction, even for dwarfs in the LG periphery (LGP), defined as those at distances $1.25<d_{\rm MW}/$Mpc$<3$ from the MW. The ``Hubble flow'' in the periphery is also affected; at fixed distance from the MW the mean recession speed, $\langle V_{\rm rad} \rangle$, varies with angular distance to M31, peaking in the anti-M31 direction and reaching a minimum behind M31. The combined M31-MW mass decelerates the local expansion; the LG ``turnaround radius'' (i.e., where $\langle V_{\rm rad} \rangle =0$) in APOSTLE is located at $r \sim 1.25$ Mpc from the LG barycentre and the pure Hubble flow (i.e., where $\langle V_{\rm rad} \rangle$ becomes comparable to $H_0*d$) is not reached out to at least $r\sim 3$ Mpc. The predicted flow is very cold, with a barycentric dispersion of less than $\sim 40$ km/s. A comparison of these predicted features with existing observations gives mixed results. There is clear observational evidence for an angular anisotropy in recession velocities around the LGP, but there is little evidence  in the spatial distribution of LGP dwarfs for a preferred direction along the MW-M31 direction. The observed local Hubble flow is also somewhat peculiar. Although the ``coldness'' of the flow is consistent with the simulations, it is significantly less decelerated:  relative to the MW, on average, all galaxies beyond $d_{\rm MW} \sim 1.25$ Mpc seem to be already on a pure Hubble flow. We argue that these oddities may result at least in part from incompleteness and inhomogeneous sky coverage, but a full explanation may need to await the completion of deep all-sky surveys able to fill the gaps in our current inventory of nearby dwarfs.
\end{abstract}


%%%%%%%%%%%%%%%%%%%%%%%%%%%%%%%%%%%%%%%%%%%%%%%%%%

%%%%%%%%%%%%%%%%% BODY OF PAPER %%%%%%%%%%%%%%%%%%

\section{Introduction}\label{sec:intro}

In the Lambda Cold Dark Matter cosmogony (LCDM), the current paradigm for structure formation, the Local Group (LG) of galaxies is thought to arise as two relatively isolated massive dark matter halos, the hosts of the Milky Way (MW) and the Andromeda (M31) galaxies, detach from the universal expansion under the influence of their mutual gravity, turn around, and start heading towards each other on a nearly radial orbit. Achieving, on first approach, the observed relative radial velocity ($\sim -109$ km/s) at the current MW-M31 separation ($\sim 774$ kpc) in roughly $14$ Gyr (the age of the Universe) suggests that the MW-M31 system turned around a few Gyrs ago after reaching a maximum separation of $\sim 1.1$ Mpc \citep{Fattahi2016}, and that their combined mass is at least a few times $10^{12}\, M_\odot$ \citep[the ``timing argument'', e.g., ][]{KahnWoltjer1959,LiWhite2008}.

Although the combined mass estimate is in reasonable agreement with current estimates of the MW and M31 virial\footnote{We define the virial boundary of a system as the radius where the mean enclosed density is 200$\times$ the critical density for closure, and refer to virial quantities with the subscript `200'.} masses \citep[see, e.g.,][and references therein]{Cautun2020,Patel2023}, other features of the LG formation scenario outlined above are seemingly at odds with LG observations.

For example, relative to the Milky Way, the current LG turnaround radius (i.e., where the mean recession velocity is $\langle V_{\rm rad}\rangle \approx 0$) should be substantially farther than $1.1$ Mpc, the expected turnaround radius of M31. \citet{Fattahi2016}, for example, estimate that the turnaround radius at present could be as large as $1.7$ Mpc from the MW. Galaxies just inside that radius should have already turned around, and have today mainly negative radial velocities. Just outside turnaround, on the other hand, galaxies should still be receding on average, but with a substantially decelerated Hubble flow. These two robust predictions are apparently in contrast with observations: all known dwarf galaxies beyond $d_{\rm MW}\sim 1.25$ Mpc from the Milky Way are receding from us, following an apparently undecelerated, ``pure'' Hubble flow.

Before analyzing this further, we note that the true boundaries of the Local Group are somewhat ill-defined. A common definition, which we follow here, defines LG galaxies as those within the current turnaround radius, estimated empirically at around $\sim 1$ Mpc from the MW-M31 barycenter \citep[see; e.g.,][]{McConnachie2012}. Galaxies just outside this radius, although receding from the LG barycenter, may still be bound to the LG. We shall hereafter refer to galaxies in the LG periphery, i.e., those at $1.25<d_{\rm MW}/$Mpc$<3$, as ``LGP dwarfs''.

Returning to the apparent disagreements mentioned above, one reason may be incompleteness in the inventory of dwarfs in the LG and its periphery, as well as their patchy distribution across the volume. Indeed, our inventory of nearby dwarfs is likely woefully incomplete, as discussed by \citet{Fattahi2020}, who conclude that as many as $\sim 50$ dwarfs as massive as the Draco dwarf spheroidal could be missing from our current inventory of  LG and  LGP members.

This incompleteness should be taken carefully into account when examining the Hubble flow around the Milky Way, as well as its dispersion. Because the Milky Way is offset from the LG barycentre, recession velocities at given distance are expected to depend on sky position, reaching a maximum in the anti-M31 direction and a minimum behind M31. (This assumes that the LG effect on the local recession velocity field is more or less symmetric relative to the LG barycentre, which should lie somewhere midway between MW and M31.)

Of all $31$ known LGP dwarfs only $2$ are within $45$ deg of M31 and $4$ in the opposite anti-M31 direction.  Such patchy coverage may therefore have a strong effect not only on how decelerated the local Hubble flow may appear, but also on estimates of its dynamical ``coldness'', an issue that has been discussed quite extensively in the literature, with conflicting claims \citep[][and references therein]{Sandage1972,Schlegel1994,Maccio2005}.

Cosmological simulations that capture the particular dynamical configuration of the MW and M31 may also offer guidance regarding where LGP dwarfs missing from our current inventory might be located.  \citet{Fattahi2020}, for example, noted that many of them should be located behind and around M31.  This anisotropy suggests that distant dwarf galaxy searches may be substantially more fruitful in some regions of the sky relative to others. Indeed, prior work has suggested that LGP dwarfs should be preferentially aligned with the MW–M31 axis, where the effects of the quadrupole of the mass distribution are maximized \citep{Penarrubia2014}.

Such guidance has already proven useful in the case of the search for new MW satellites in the vicinity of the Magellanic Clouds. Indeed, \citet{Sales2011} used cosmological N-body simulations to predict that the surroundings of the Clouds should ``prove a fertile hunting ground for faint, previously unnoticed MW satellites'', a prediction that became spectacularly true with the discovery of a number of nearby dwarfs in the DES survey, one of the first to target large fractions of the southern sky to magnitudes deep enough to allow for the identification of new MW satellites and LG members \citep{Bechtol2015}.

We revisit these issues here, using simulations from the APOSTLE\footnote{APOSTLE stands for ``A Project Of Simulating The Local Environment'' \citep{Sawala2016,Fattahi2016}.}  project. Our main goal is to characterize expected anisotropies in the spatial distribution of dwarfs in the Local Group, with particular emphasis on the spatial and kinematic properties of the LGP dwarf population, and to compare them with current observations.

We compare simulations and observations mainly in the MW-centric frame, since the lack of proper motions for most nearby dwarfs means that it is not possible to transform accurately their velocities to an LG-centric frame. This approach differs from that commonly adopted in earlier work, where LG-centric velocities are estimated by simply projecting the observed radial velocities onto a frame where the dispersion in the Hubble flow of distant galaxies is minimized \citep[see; e.g.,][and references therein]{Karachentsev2009,Penarrubia2014}. This approach may introduce biases not only  because it neglects tangential velocities, but also because of the aforementioned lack of homogeneous spatial coverage in our current  inventory of nearby dwarfs, which may compromise the minimization procedure.

This paper is organized as follows. Sec.~\ref{sec:methods} introduces the numerical simulations and observational dataset of 
galaxies used. Sec.~\ref{sec:disLGdwarfs} quantifies the anisotropies in the spatial distribution of 
 dwarfs in the APOSTLE simulations, while Sec.~\ref{sec:hubble} examines anisotropies in the recession speeds of  dwarfs, with emphasis on the deceleration and dispersion of the local Hubble flow. Finally, we summarize our results and discuss their implications in Sec.~\ref{sec:conclu}.

\begin{figure*}
\centering
 \ifembedvideo
 \embedvideo*{\includegraphics[width=\textwidth]{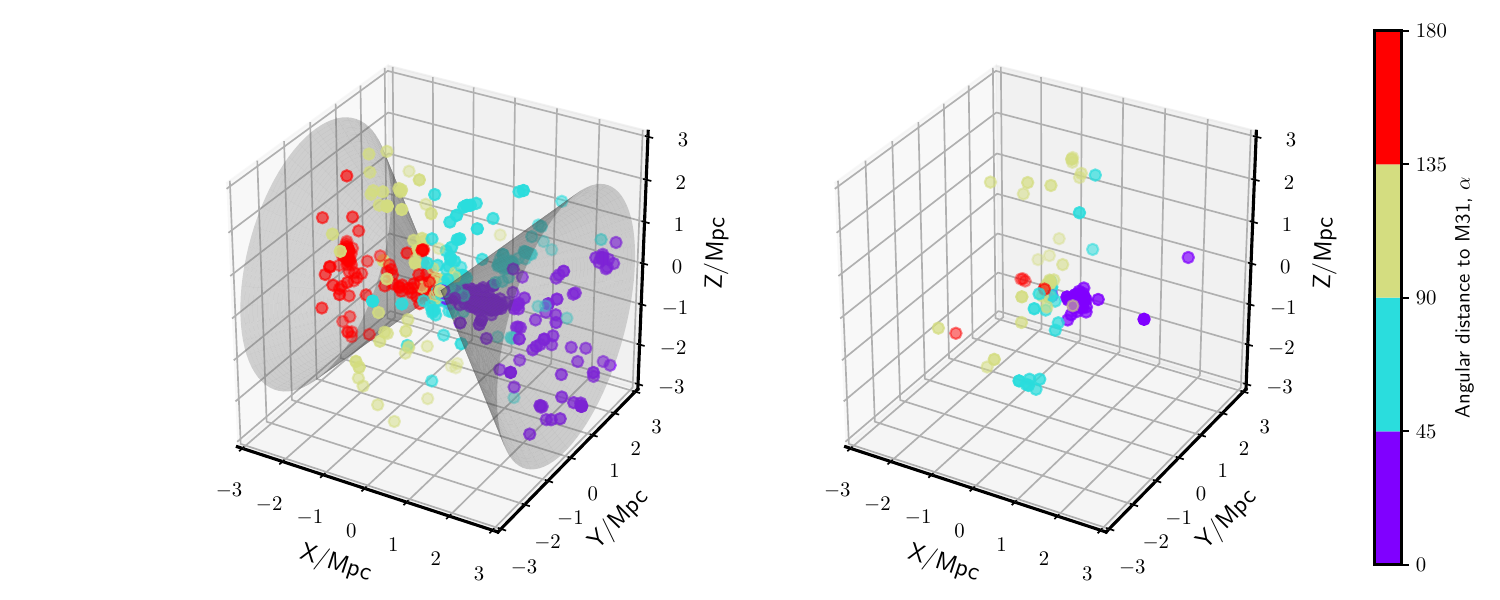}}{figures/video3DLG_longfast.mp4}
 \else
\includegraphics[width=\linewidth]{figures/3Dproj_SimsObs_new.pdf}
\fi
\caption{3D positions of $M_*>10^5\, M_\odot$ Local Group dwarf galaxies within a radius of $\sim3.5$ Mpc  from the LG midpoint.  \textit{Left:} APOSTLE, centered on the MW analog: To enhance visually the shape of the spatial distribution, we stack the data from the 4 APOSTLE-L1 volumes.
\textit{Right:} Observational data, centered on the MW.  
In both panels, the X-axis is aligned with the MW-M31 direction.  Galaxies are colored according to $\alpha$, the angular distance between a system and the direction to M31; see colorbar. Gray spherical cones delineate an aperture of $45^\circ$ around this axis for reference.
}
\label{fig:3d}
\end{figure*}

\section{Methods}\label{sec:methods}
\subsection{Numerical simulations} \label{sec:sims} We use the APOSTLE project, a suite of ``zoom-in'' cosmological hydrodynamical simulations of LG-like environments that include two primary halos with masses, relative distances, and relative radial and tangential velocities chosen to be roughly consistent with current observational constraints for the MW-M31 pair \citep{Fattahi2016}.

In this work we have used the $z=0$ outputs of 4 different APOSTLE volumes from the highest ``L1'' resolution level, with initial dark matter and gas particle masses of $m_{\rm dm}\sim5\times10^{4}$ M$_\odot$ and $m_{\rm gas}\sim1\times10^{4}$ M$_\odot$, respectively, and a gravitational softening length of $134$ pc at $z=0$. The average combined virial mass of the MW and M31 pair is $2.7\times 10^{12}$ M$_\odot$, with mass ratios in the range $0.65$-$0.97$.
In each APOSTLE volume, the zoomed-in region includes a sphere of radius $r\sim3.5$ Mpc around the midpoint of the MW-M31 pair, which is fully enclosed within the highest resolution volume.

APOSTLE has been run with the EAGLE galaxy formation code \citep{Schaye2015,Crain2015}, which includes subgrid physics prescriptions for radiative cooling of gas,  star formation in gas particles exceeding a metallicity-dependent density threshold,   stellar feedback in the form of stellar winds, radiation pressure and supernovae, as well as an homogeneous X-ray/UV background radiation. The model also accounts for supermassive black hole growth and AGN feedback, but we note that these have negligible effects in the APOSTLE volume as it is dominated by low-mass, dwarf galaxies.

The APOSTLE project adopts a flat $\Lambda$CDM cosmological model with WMAP-7 parameters \citep{Komatsu2011}: $\Omega_{\rm m}=0.272$; $\Omega_{\Lambda} = 0.728$; $\Omega_{\rm bar} = 0.0455$; $H_0 = 100\, h$ km s$^{-1}$ Mpc$^{-1}$; $\sigma = 0.81$; $h = 0.704$.

\subsubsection{Simulated galaxies}
APOSTLE halos were identified using the friends-of-friends (FoF) groupfinding algorithm \citep{Davis1985} assuming a linking length of 0.2 times the mean interparticle separation.  Self-bound substructures within FoF groups were then identified using  SUBFIND \citep{Springel2001}.

Luminous galaxies form in APOSTLE at the centre of halos that exceed a redshift-dependent ``critical mass'', set by the UV-ionizing background \citep{BenitezLlambay2020}. At $z=0$ this threshold corresponds to a virial mass of  $M_{200}\sim 10^{9}$ M$_\odot$, resolved with $>2\times10^4$ dark matter particles in APOSTLE-L1 \citep{PereiraWilson2023}.

Systems with lower-than-critical mass that host a luminous galaxy are either halos which were over the critical boundary in the past but whose recent mass accretion history has been uncharacteristically slow, or the result of tidal stripping, which may reduce the total dark matter mass of a system that orbits a more massive host.

In this work we are mainly interested in 
%LG
nearby  ``field'' dwarfs that are not satellites of either the MW or M31; i.e., those found at $z=0$ outside the virial radii of the two massive APOSTLE primaries, and within a sphere of radius $r \sim 3.5$ Mpc from the MW-M31 midpoint.

We shall hereafter refer to luminous galaxies within the virial radius of either primary as ``satellites''.
To minimize numerical resolution effects, we shall only use for our analysis galaxies with at least $\sim 10$ star particles, or stellar masses $M_{*}>10^5\, M_\odot$.

\subsection{Observational data} \label{sec:data}

In this work we consider all currently known dwarf galaxies within $3$ Mpc of the midpoint between the MW and M31. We use position (RA, dec), distance modulus $(m-M)$ and line-of-sight velocity data from the latest update of \citet{McConnachie2012}'s Nearby Dwarf Galaxy Database\footnote{See \url{https://www.cadc-ccda.hia-iha.nrc-cnrc.gc.ca/en/community/nearby/}, and references therein.}.

We consider systems within 300 kpc of the MW or M31 as ``satellites'' of that primary, and those further away as ``field'' dwarfs. Our total sample consists of $142$ dwarfs of which $48$ are field galaxies. To be consistent with the simulation limitations, we also impose a minimum stellar mass of $M_*=10^5\, M_\odot$. This cut removes from our sample only MW satellites, which are not the main focus of our study, leaving a total of $96$ dwarfs.

From the catalogued data, we have computed Galactocentric positions and radial velocities assuming a Galactocentric distance for the Sun of $R_\odot=8.29$ kpc, a circular velocity for the local standard of rest (LSR) of $V_0=239$ km/s \citep{McMillan2011}, and a peculiar velocity with respect to the LSR of $(U_\odot,V_\odot,W_\odot) = (11.1, 12.24,7.25)$ km/s \citep{Schonrich2010}.

\begin{figure*}
\centering
\includegraphics[width=0.8\linewidth]{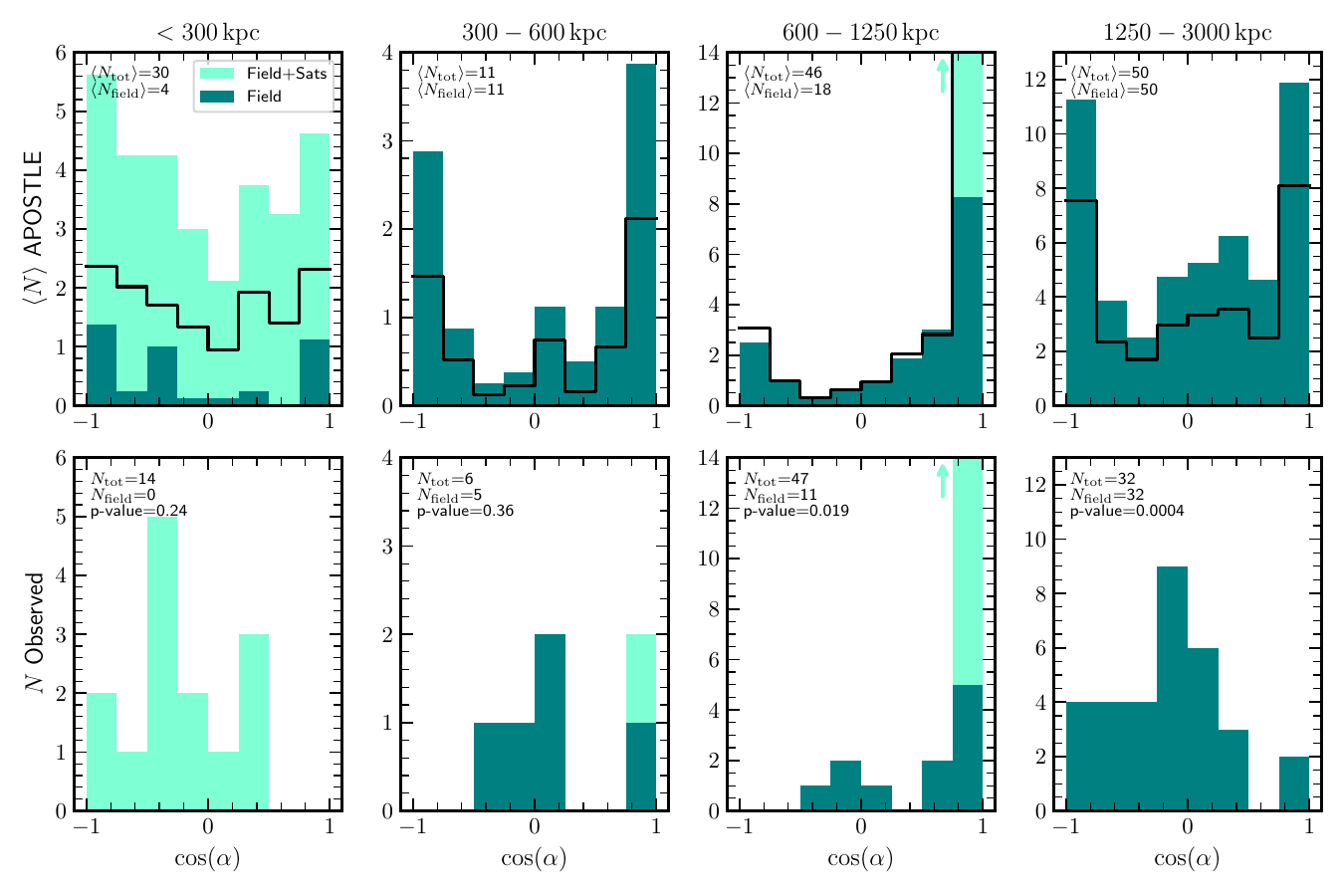}
\caption{Distribution of the (cosine of) angular distances between dwarf galaxies and the direction to the M31 analog, in concentric spherical shells with Galactocentric distances as indicated in the upper legend of each panel.  Light green histograms show \textit{all} galaxies (field + satellites) while teal histograms show only field galaxies.  The number of dwarfs considered in each spherical  shell is quoted in the panels.  
\textit{Top:} APOSTLE dwarf galaxies.  We indicate average number of galaxies,  i.e., total normalized by 8, as we consider 8 different LG configurations alternating the MW and M31 identification (see text). The black histogram indicates the result of using, on average, only as many systems as are actually observed (bottom panels) in each spherical shell. \textit{Bottom:} Observational data.  The first shell includes only MW satellites, whereas the third shell includes mainly M31 satellites.  The simulations show an excess of objects along the MW-M31 direction (at $\cos(\alpha)\approx 1$ and $-1$); a trend that is not readily seen in the observational data.  }
\label{fig:hist}
\end{figure*}

\begin{figure*}
  \centering
  \includegraphics[width=1\linewidth]{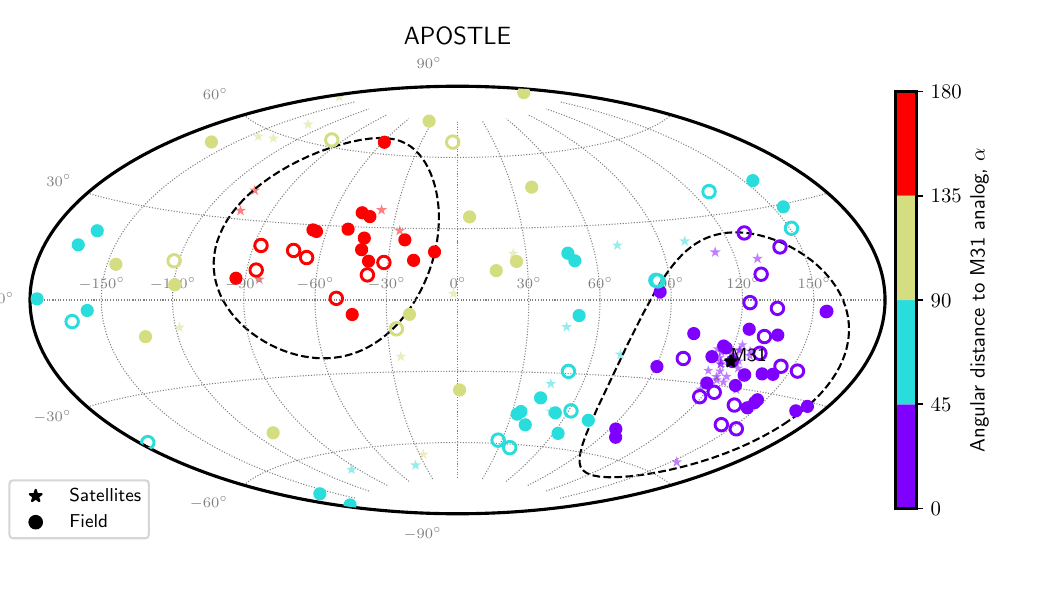}\\ %0.95
  \includegraphics[width=1\linewidth]{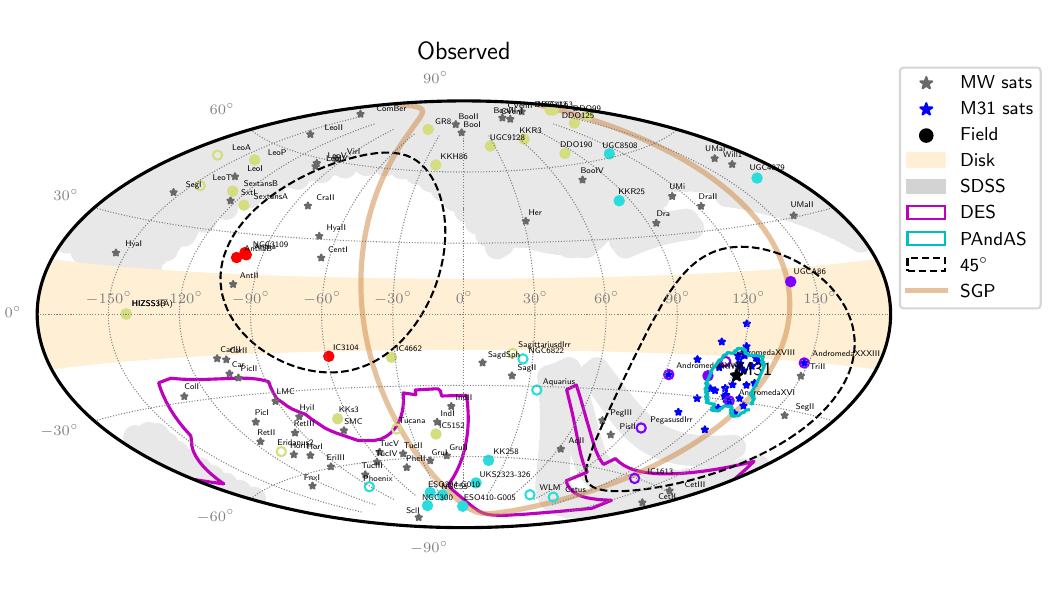}
  \caption{Aitoff projections of nearby dwarf galaxies in Galactic coordinates.  \textit{Top:} One eighth of randomly selected APOSTLE galaxies are shown after stacking 8 different configurations, rotated so in each the direction to M31 matches the observed position of M31 in the sky (see text). \textit{Bottom:} Observed LG and periphery.  Satellite galaxies are shown with star symbols (gray for MW satellites and blue for M31 satellites). Field dwarfs are shown as circles colored according to $\alpha$. Filled circles correspond to LGP dwarfs  and open circles to the rest of field dwarfs at distances $<1.25$ Mpc. Black-dashed circles mark an area of $45^\circ$ around the MW-M31 direction.  In the bottom panel, patches of different colors illustrate the footprints of the observational surveys indicated in the legend. The effect of Galactic disk obscuration is shown in yellow, and the Supergalactic Plane is marked with a brown line.  }
\label{fig:aitoff}
\end{figure*}

\begin{figure*}
\centering
\includegraphics[width=0.75\linewidth]{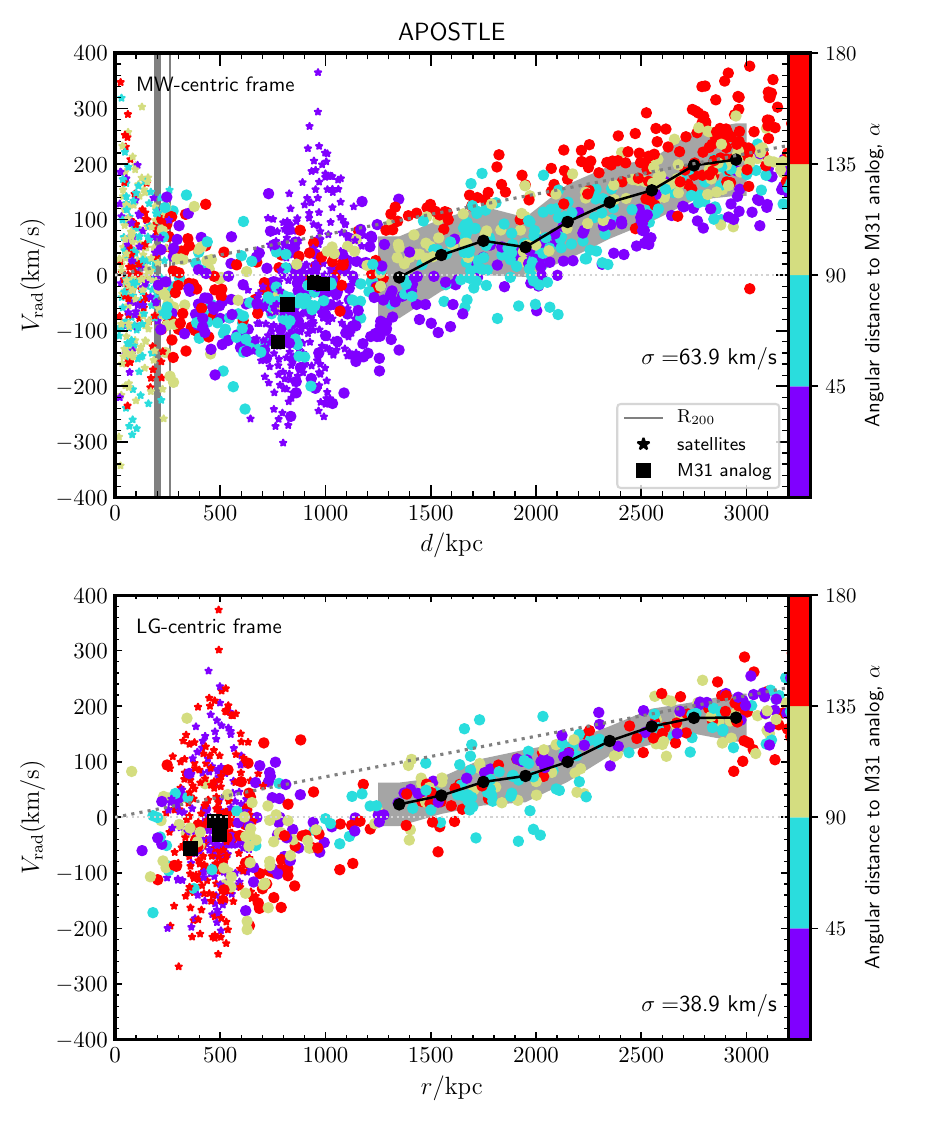}
\caption{Radial velocity versus distance for nearby dwarf galaxies in the APOSTLE simulations.
We stack  8 different LG configurations and plot only dwarfs with $M*>10^5$ M$_\odot$ within 3.5 Mpc of the LG midpoint.
\textit{Top:} Centered on each of the 8 APOSTLE LG primaries.  
\textit{Bottom:} Centered on the barycenter of each LG. % (4 configurations).
Satellites are shown with star symbols while field galaxies are shown as circles. All dwarfs are colored according to  $\alpha$ as in previous figures (see legend).
Black circles indicate the mean radial velocity computed in radial bins within $1.25<d_{\rm MW}$/Mpc$<3$, and the gray shaded areas show the $\pm 1\sigma$ deviation from the mean. The gray dotted line shows a pure Hubble law with $H_0=73$ km/s/Mpc \citep{Riess2022}.  
Tab.~\ref{tab:H} gives linear fits to the
recession velocity of LGP APOSTLE dwarfs in the MW-centric reference frame (upper panel), expressed in the form $V_{\rm rad}=\mathcal{H} (d_{\rm MW}/$Mpc$-1.25)+ V_{1.25}$, for each angular bin in $\alpha$.
}
\label{fig:Hubb}
\end{figure*}

\section{Results}\label{sec:results}
\subsection{The spatial distribution of Local Group dwarf galaxies} \label{sec:disLGdwarfs}

The left panel of Fig.~\ref{fig:3d} shows the 3D-positions of APOSTLE nearby dwarfs (field and satellite galaxies) with stellar masses $M_{*}>10^5$ M$_\odot$.  The coordinate system is centered on the smaller  of the two primaries in each volume (we shall refer to it as the ``MW analog''\footnote{In APOSTLE the less massive of the two halo primaries is usually considered the Milky Way, although for some of the analysis we shall drop the distinction between ``MW'' and ``M31'' analogs.}),   with the ``$X$'' axis being coincident with the direction connecting the two primaries. (M31 is thus located on the $X$-axis at roughly $X\sim 800$ kpc.) The plot includes all simulated dwarfs within  a radius of $3.5$ Mpc  from the LG midpoint of each of the 4 APOSTLE-L1 volumes, stacked together (i.e., the plot shows four times as many dwarfs as would be expected in the true LG; the increased number helps to visualize the 3D structure of the LG).  

Galaxies are colored by the angle, $\alpha$, between the position of a galaxy and the direction to the M31 analog; i.e.,  the angular distance between the dwarf and the MW-M31 axis. Systems colored purple are those closer than $45^\circ$ to the direction towards M31, while objects that are colored red are those ``behind'' the MW  along the same axis ($135^\circ<\alpha<180^\circ$). For reference, the  gray spherical cones in Fig.~\ref{fig:3d} highlight the $\alpha=45^\circ$ and $\alpha=135^\circ$ boundaries in the volume.

The spatial distribution of APOSTLE nearby dwarfs is not isotropic, but clearly elongated, with the majority of objects located close to the $X$-axis joining the MW and M31 analogs. This seems true at all radii, not only for relatively nearby galaxies, which include the satellites of the main halos, but also for the more distant LGP dwarfs far outside their virial radii. 

The anisotropic spatial distribution of APOSTLE nearby dwarfs is quantified in Fig.~\ref{fig:hist}, where we plot the angular distribution of galaxies on the sky in four different spherical shells centred on the MW analog.  Each panel shows a histogram of $\cos(\alpha)$ for all $M_*>10^5\, M_\odot$ galaxies in a shell.

The top row in Fig.~\ref{fig:hist} corresponds to the average of $8$ APOSTLE configurations, two for each APOSTLE volume, alternating the designation of MW or M31 analogs. This helps to reduce noise and to characterize more robustly the anisotropic distribution in each shell.

Light-green histograms in Fig.~\ref{fig:hist} show results for all galaxies (i.e., satellites + field), while teal histograms show results  considering only field dwarfs. The first radial bin (shown in the leftmost panels) is, as expected, dominated by the satellite population of each primary chosen as centre.

The second radial bin ($300<d_{\rm MW}$/kpc$<600$) is chosen to exclude most satellites, whereas the third bin includes the majority of the satellites of the second primary. Note as well that the second bin  includes a high fraction of so-called ``backsplash'' dwarfs ($>50\%$ within $2\times r_{200}$),  i.e.,  galaxies found presently outside the virial radii of a primary but which were in the past inside $r_{200}$ \citep{SantosSantos2023b}. The final, outermost radial bin (rightmost panels in Fig.~\ref{fig:hist},  $1.25<d_{\rm MW}$/Mpc$<3$) includes only LGP field dwarfs.

A spatially uniform galaxy distribution would show in Fig.~\ref{fig:hist} as a constant number of systems as a function of $\cos(\alpha)$. This is clearly not the case for APOSTLE dwarfs, which show a clear excess at $\cos(\alpha)\approx 1$ and $\cos(\alpha)\approx -1$ for all radial shells. The ``U-shaped'' distributions indicate a clear preference for dwarfs to align with the MW-M31 axis, a preference which persists even for LGP dwarfs located as far as $3$ Mpc away from either primary.

Is the same predicted anisotropy observed in our Local Group? The right-hand panel of Fig.~\ref{fig:3d} shows the positions of observed  dwarfs in a reference frame with the Milky Way at the origin. A total of $96$ galaxies with $M_*>10^5\, M_\odot$ are known within $3$ Mpc from the Milky Way, whereas, on average, each of the APOSTLE volumes has $176$ such galaxies within the same volume. In other words, APOSTLE predicts that roughly $80$ galaxies as massive as $M_*>10^5\, M_\odot$ may be currently missing from our inventory of nearby dwarfs \citep[see also][]{Fattahi2020}. 
Many of these missing dwarfs are predicted to be in the outermost radial shell presented in Fig.~\ref{fig:hist}. In this figure, the number of such galaxies observed in each shell is listed in each panel of the bottom row of that figure, and may be compared with the average number of APOSTLE dwarfs quoted in the top row.

Intriguingly, the observed dwarfs in the LG and periphery do not seem to follow the same anisotropic distribution predicted by APOSTLE. Indeed, there is little evidence for a preferred alignment of dwarfs with the axis defined by M31 and the MW; if anything, the opposite trend appears to prevail, with clear peaks at $\cos(\alpha)\approx 0$, both in the satellites of the Milky Way (bottom left panel of Fig.~\ref{fig:hist}) and in the LGP dwarfs (bottom-right panel of Fig.~\ref{fig:hist}).

Could the difference be due to small-number statistics, or caused by incompleteness in our inventory of currently known nearby dwarfs? The former seems unlikely, as shown by the solid black histograms in the upper panels of Fig.~\ref{fig:hist}, which indicate the result of choosing at random in each APOSTLE radial bin only as many galaxies as are available in the observational sample. The ``U-shaped'' anisotropy is still clearly noticeable for APOSTLE galaxies but not in the observations.  A K-S test rejects the hypothesis that the distributions in the upper and lower panels of Fig.~\ref{fig:hist} are drawn from the same parent distribution at high confidence in all cases where numbers permit  (see p-values in the legends).

If the lack of ``U-shaped'' anisotropy in the observed dwarfs is caused by incompleteness, then there should be a number of undetected dwarfs both behind M31 and in the anti-M31 direction.  We show this in Fig.~\ref{fig:aitoff}, where the colored circles in the upper Aitoff diagram shows the on-sky distribution of
APOSTLE field dwarfs randomly sampled from 8 possible LG configurations, obtained by alternating the MW and M31 analogs.
Filled circles represent LGP field dwarfs while open circles indicate the rest of field dwarfs at distances $<1.25$ Mpc from the primary.
The satellites of each primary are shown by stars 
but are de-emphasized in this plot.

The high density of red and purple LG dwarfs is quite clear in APOSTLE, compared to that of galaxies further away from the MW-M31 direction: indeed, while the circles around M31 and the anti-M31 direction cover only $30\%$ of the sky, they contain roughly $57\%$ of all dwarfs within $3.5$ Mpc from the LG barycenter (or $54\%$ of all dwarfs in the LGP).  In terms of numbers, according to APOSTLE we would expect to find, on average, $8$ LG and $13$ LGP dwarfs in the red circle, and $14$ LG and $25$ LGP dwarfs in the purple one, and a total of about $12$ LG and $33$ LGP dwarfs in other regions of the sky (note this count is for dwarfs within $3.5$ Mpc from the LG barycentre).

In contrast, there are very few {\it observed} LGP dwarfs in the red and purple areas, as shown in the bottom panel of Fig.~\ref{fig:aitoff}: only $2$ in the M31 direction, and $4$ in the anti-M31 direction. One reason for this may be that imaging surveys have covered the sky unevenly, reducing the possibility of detecting dwarfs in the M31 and anti-M31 directions. The ``zone of avoidance'' (i.e., $b<15^\circ$) caused by Galactic disk obscuration may also play a role: given M31's low Galactic latitude, roughly $\sim 39\%$ of the sky with $\alpha<45^\circ$ or $\alpha>135^\circ$ is obscured by the disk.

The bottom panel of Fig.~\ref{fig:aitoff} also suggests that the anti-M31 direction, in particular, might not have been surveyed as deeply as other parts of the sky.  Indeed, most LGP dwarfs have been found by visual inspection of photographic plates \citep[see; e.g.,][]{Whiting1999,Karachentseva1998}, which suggests that there is scope for new discoveries with the advent of digital surveys of the whole sky. The anti-M31 direction is also outside the footprint of both the Sloan Digital Sky Survey, which covers most of the northern sky \citep[SDSS; gray-shaded area]{Willman2005}, and of the DES survey \citep[purple contour,][]{Drlica-Wagner2015}, which has imaged a large fraction of the southern sky. There have been several surveys in the general direction of M31 \citep[most notably the PAndAS survey, shown with a cyan contour,][]{McConnachie2009} but they only cover a very small fraction of the $\alpha<45^\circ$ area of the sky around M31.

Recently, \citet{McNanna2023} have searched for field dwarfs in the DES footprint with distances between $300$ kpc and $2$ Mpc from the MW. Although their search should detect all $M_* > 10^5\, M_\odot$ galaxies like the ones we study here, they report no new discoveries aside from the $7$ already known dwarfs in that volume. For comparison, averaging all volumes and various orientations, APOSTLE predicts a median number of $3$ in that region, with a $10$th and $90$th percentile of $0$ and $8$, respectively (and a full range that goes from $0$ to $37$ dwarfs). Given the large volume-to-volume scatter in APOSTLE, we find no obvious conflict between the result of the latest search in the DES footprint and the results from the simulations.

To summarize, the APOSTLE runs predict a clear anisotropic distribution for nearby galaxies, which are expected to align strongly with the M31 and anti-M31 directions. This prediction seems at odds with currently available data on LG and LGP dwarfs. The reason for the disagreement is unclear, but, if due to incompleteness, then the simulations suggest that those directions could prove fruitful targets for future searches of LGP dwarfs.

\begin{table}
	\centering
	\caption{Parameters of linear fits to the recession velocity of LGP dwarfs in APOSTLE (in the MW-centric reference frame) of the form 
	$V_{\rm rad}=\mathcal{H} (d_{\rm MW}/$Mpc$-1.25)+ V_{1.25}$, 
	for each of the angular bins $\alpha$.  These are illustrated as colored lines in the inset to Fig.~\ref{fig:Hubbobs}.}
	\label{tab:H}
			\begin{tabular}{l l l}   %APOSTLE
		\hline
		 APOSTLE  &  $\mathcal{H}$ [km/s] & $V_{1.25}$ [km/s]  \\
		\hline
		\hline
	    Mean MW-centric & 134.3 &  -17.3 \\
		$\alpha<45^{\circ}$ & 120  &  -61.2 \\
		$45^{\circ}>\alpha<90^{\circ}$ & 102.4 & -13.4 \\
		$90^{\circ}>\alpha<135^{\circ}$ & 104.6  & 26.5 \\
		$\alpha>135^{\circ}$ & 97.3 & 86.0 \\
		\hline
		Mean LG-centric & 108.3 & 7.8 \\
		\hline
	\end{tabular}
	
\end{table}

\subsection{The Local Hubble Flow in APOSTLE}\label{sec:hubble}

The anisotropic distribution of mass predicted by APOSTLE should also have consequences on the velocity field of galaxies around the Milky Way. As discussed in Sec.~\ref{sec:intro}, galaxies beyond the LG turnaround radius should be expanding away with a decelerated Hubble flow. The velocity field is also expected to be fairly symmetric relative to the LG barycentre, and to show a clear asymmetry when expressed in the Galactocentric (MW) reference frame. From that perspective, distant galaxies at given distance from the MW, $d_{\rm MW}$, should have recession velocities which depend on the angular distance to the direction of M31, peaking in the anti-M31 direction and having a minimum behind M31.

We show this in the upper panel of Fig.~\ref{fig:Hubb}, which shows the radial velocity relative to the MW analog in APOSTLE, as a function of MW-centric distance. Satellite galaxies are shown with star symbols, while field galaxies are shown as circles, all  colored according to  $\alpha$ as in previous figures. Black squares indicate the radial distance and velocity of the ``M31 analog'' in each volume,  and thin gray lines indicate the virial radii of all primaries, for reference. 
Results are stacked for 8 LG configurations, obtained by  alternating  the MW and M31 analogs.

Galaxies in Fig.~\ref{fig:Hubb} are, as in previous plots, colored by their angular distance to the M31 analog. In particular, systems in purple are in the general direction of M31, and those in red are located close to the anti-M31 direction. The black points and gray shade represent the mean and $\pm 1\sigma$ standard deviation of $V_{\rm rad}$ computed in radial bins  in the range $1.25$ to $3$ Mpc. The average rms about the mean velocity  is quoted in each panel.

As may be seen from the top panel of Fig.~\ref{fig:Hubb}, the recession velocities of LGP dwarfs, as seen from the MW, are clearly modulated by their angular distance to M31. The difference is not subtle.  Behind M31, galaxies at $d_{\rm MW} \sim 2$ Mpc have a mean recession velocity of just $\sim 30$ km/s. At the same distance, galaxies in the anti-M31 direction are receding on average at roughly $\sim 160$ km/s.

This angular dependence hinders a proper characterization of the local recession velocity field, including a precise determination of the LG turnaround radius (where the mean radial velocity vanishes), the dispersion about the mean flow (the ``coldness'' of the local Hubble flow), and how decelerated the local velocity field is relative to the pure Hubble flow.

As expected, the angular dependence disappears when referring velocities and distances to the barycentre of the M31-MW system, as shown in the bottom panel of Fig.~\ref{fig:Hubb}. This makes it easy to estimate the average APOSTLE LG turnaround radius ($r_{\rm ta}\sim 1.2$ Mpc from the LG barycentre), and the velocity dispersion about the mean flow ($\sim 40$ km/s). The flow is clearly decelerated relative to a pure Hubble flow (indicated by the gray dotted line in the bottom panel of Fig.~\ref{fig:Hubb}): even galaxies as far away as $\sim 3$ Mpc from the LG barycentre have not yet reached the pure Hubble flow.

\begin{figure*}
\centering
\includegraphics[width=0.85\linewidth]{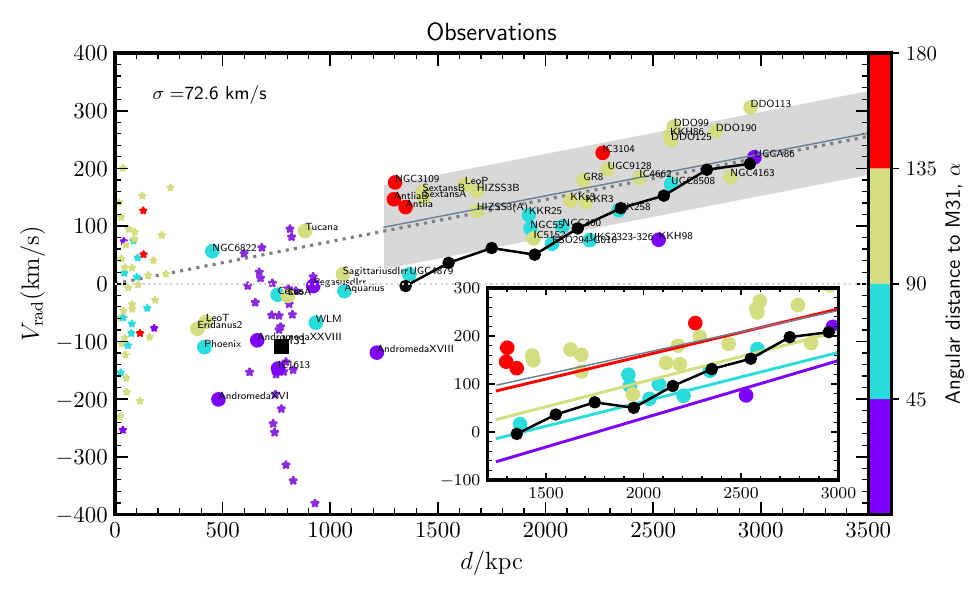}
\caption{Galactocentric radial velocity versus radial distance for observed dwarf galaxies.  Symbols and color-coding are the same as in Fig.~\ref{fig:Hubb}.  A thin gray line and shaded area indicate the linear fit and standard deviation obtained for galaxies in the range $1.25<d_{\rm MW}$/Mpc$<3$.  For comparison,  we overplot the black circles  shown in the top panel of Fig.~\ref{fig:Hubb}, i.e.,  the mean distance-velocity relation for APOSTLE. The dotted line marks a pure Hubble law with $H_0=73$ km/s/Mpc. The inset panel shows the same observational data with the addition of 4 lines representing linear fits to APOSTLE LGP dwarfs in each of the four $\alpha$  bins (see Tab.~\ref{tab:H}).
}
\label{fig:Hubbobs}
\end{figure*}

How do these results compare with observational data for the Local Group? In this case, we can only compute the Galactocentric flow, because proper motions for all galaxies, which are unavailable, or a detailed 3D velocity model, would be needed to refer recession velocities to the LG barycentre \citep[see; e.g.,][]{Karachentsev2009,Penarrubia2014}.

Fig.~\ref{fig:Hubbobs} shows the distance-recession velocity relation for known dwarfs, measured with respect to the MW.  The 
thin solid gray line indicates a linear fit to the data in the distance range $1.25<d_{\rm MW}$/Mpc$<3$. The shaded area around that fit shows the corresponding $1\sigma$ standard deviation, of order $\sim 65$ km/s. Note that this is in excellent agreement with the coldness of the Galactocentric local Hubble flow in APOSTLE (see upper panel of Fig.~\ref{fig:Hubb}), which implies that the APOSTLE runs have no problem accounting for the observed coldness of the local Hubble flow.

There is also good qualitative agreement between APOSTLE and observations regarding the angular dependence of the recession velocity. As visual inspection shows, at given distance, the recession velocities are highest in the anti-M31 direction and lowest behind M31, with a velocity difference between antipodal directions exceeding $100$ km/s. We note that the reflex motion of the Milky Way caused by the recent infall of the Magellanic Clouds system could add a similar dipole-like effect along a similar axis, but the amplitude in that case is likely to be much smaller, roughly of only $\sim 30$ km/s according to \citet{Petersen2020}. The velocity anisotropy in the MW frame is therefore mainly due to the offset between the Milky Way and the LG barycentre, as shown for APOSTLE in Fig.~\ref{fig:Hubb}.

The main difference between observations and simulations is that the observed local Hubble flow seems much less decelerated than that of APOSTLE. At given distance form the MW, APOSTLE's Galactocentric recession velocities (shown by the black connected circles in Fig. ~\ref{fig:Hubbobs}) are well below observed ones (shown by the solid gray line; the dotted gray line indicates a pure Hubble flow with $H_0=73$ km/s/Mpc, \citealt{Riess2022}).

One way of reconciling this difference would be to assume that the combined MW+M31 mass is much lower than assumed in APOSTLE, but this seems unlikely. Indeed, as discussed by \citet{Fattahi2016}, a total combined mass of order  $10^{11.5}$ M$_\odot$ would be needed, at least an order of magnitude lower than current estimates. 

Could the  disagreement be due instead to the patchy sky coverage of the observational sample? We explore this in the inset to Fig.~\ref{fig:Hubbobs}, where we compare linear fits to the MW-centric APOSTLE Hubble flow on different parts of the sky (colored lines) with the observational data. The disagreement is worse for galaxies with $\alpha>90^\circ$ (i.e., the hemisphere in the anti-M31 direction, shown in red and green), where the observed recession speeds systematically exceed  the APOSTLE predictions. Although the disagreement  seems clear, it is important to keep in mind how sparse the sky coverage of LGP dwarfs is: only $4$ ``red'' distant dwarfs are known in the anti-M31 direction, and only $2$ (shown in purple) are known behind M31.

It is therefore certainly possible that the comparison may change if even a few more systems are added in each of these directions. APOSTLE predicts that about $\sim 60$ \textit{field} dwarfs with $M_*>10^5\, M_\odot$ are likely be missing from our nearby dwarf galaxy inventory, $41$ of them LGP dwarfs (see Sec.~\ref{sec:disLGdwarfs}).  In particular, the lack of deceleration in the observed MW-centric Hubble flow could be due to the paucity of LGP dwarfs behind M31 (i.e., purple circles in Fig.~\ref{fig:Hubbobs}). Finding some of the ``missing dwarfs'' predicted by APOSTLE in the MW-M31 direction could certainly impact our understanding of the Hubble flow in the outskirts of the Local Group.

If not due to incompleteness, it is possible that the apparent lack of deceleration in the velocity field of LGP dwarfs may be due to the large-scale distribution of matter around the Local Group. This could in principle be investigated using simulations tailored to reproduce not only the LG environment but also that of its surrounding large-scale structure  \citep[see; e.g.,][]{Carlesi2017,Libeskind2020,Sawala2022}.

\section{Summary and Conclusions}\label{sec:conclu}

We have used the APOSTLE suite of cosmological hydrodynamical simulations to study anisotropies in the spatial distribution and kinematics of dwarf galaxies in the Local Group and periphery, out to a distance of $\sim 3$ Mpc from the MW. The simulations show that the anisotropy induced by the presence of two massive primaries  on first approach is reflected in the spatial distribution of nearby dwarf galaxies. At all distances from the MW, the simulations predict a strong preference for dwarfs to be located close to the axis defined by the MW-M31 direction, from the satellites of the primary galaxy to even the ``distant'' LGP dwarfs, defined as those in the LG periphery, i.e., at distances $1.25<d_{\rm MW}/$Mpc$<3$ from the MW.

The local ``Hubble flow'' of the LGP dwarfs is also expected to be anisotropic if measured in the Galactocentric rest frame. At fixed distance from the MW the mean recession speed, $\langle V_{\rm rad} \rangle$, varies with angular distance to M31, peaking in the anti-M31 direction and reaching a minimum behind M31, mainly as a result of the offset between the MW and the LG barycentre, which lies somewhere between MW and M31.

The combined M31-MW mass also decelerates the local Hubble flow of LGP dwarfs; the LG ``turnaround radius'' (i.e., where $\langle V_{\rm rad} \rangle =0$) in APOSTLE is located at $r_{\rm ta} \sim 1.2$ Mpc from the LG barycentre and the pure Hubble flow (i.e., $\langle V_{\rm rad} \rangle = H_0*r$) is not reached out to at least $r\sim 3$ Mpc. The predicted flow is very cold, with a barycentric dispersion of less than $\sim 40$ km/s.

A comparison of these features with existing observations raises interesting questions. Although there is agreement with the predicted angular anisotropy in recession velocities around the Milky Way, there is little evidence  in the spatial distribution of LGP dwarfs for a preferred direction along the MW-M31 direction.

The ``coldness'' of the local Hubble flow also seems consistent with the simulations, but it is significantly less decelerated. Indeed, in the Galactocentric frame, all dwarfs beyond $r \sim 1.25$ Mpc seem to be receding with velocities consistent with a pure, undecelerated Hubble flow. Although the reason for these differences is so far unclear, 
APOSTLE also predicts that the true number of LGP dwarfs should be substantially higher than observed, suggesting that our local inventory of 
%LG
 dwarfs is rather incomplete.

Another reason for the disagreement may be that the APOSTLE volume selection made no attempt to account for structures beyond $\sim 3$ Mpc from the LG barycentre. The presence of large galaxies just outside that volume, like M81 or NGC 5128, as well as the influence of the Virgo cluster or the Local Void,  may all have an influence over the spatial distribution and velocity field of LGP dwarfs and clearly need to be taken into account in future, higher fidelity simulations of the LG volume.

It is thus possible that the oddities described above may result at least in part from incompleteness and inhomogeneous sky coverage, but a full explanation will need to await the completion of deep all-sky surveys able to fill the gaps in our current inventory of the Local Group, and of simulations able to fully reproduce the configuration of the Local Group within the larger-scale distribution of matter in the local Universe.

\section*{Acknowledgements}
ISS acknowledges support from the European Research Council (ERC) through Advanced Investigator grant to C.S. Frenk, DMIDAS (GA 786910).  We wish to acknowledge the generous contributions of all those who made possible the Virgo Consortium’s EAGLE/APOSTLE simulation projects. This work used the DiRAC@Durham facility managed by the Institute for Computational Cosmology on behalf of the STFC DiRAC HPC Facility (www.dirac.ac.uk). The equipment was funded by BEIS capital funding via STFC capital grants ST/K00042X/1, ST/P002293/1, ST/R002371/1 and ST/S002502/1, Durham University and STFC operations grant ST/R000832/1. DiRAC is part of the National e- Infrastructure.  This research made use of Astropy (http://www.astropy.org) a community-developed core Python package for Astronomy.

\section*{Data Availability}
The simulation data underlying this article can be shared on reasonable
request to the corresponding author.
The references for the observational data for nearby dwarfs used in this article are listed in Sec.~\ref{sec:data}.



\bibliographystyle{mnras}
\bibliography{archive} 

%%%%%%%%%%%%%%%%% APPENDICES %%%%%%%%%%%%%%%%%%%%%
%\appendix


% Don't change these lines
\bsp	% typesetting comment
\label{lastpage}
\end{document}

%% file: embed_video.tex
\usepackage[bigfiles]{pdfbase}
\ExplSyntaxOn
\NewDocumentCommand\embedvideo{smm}{
  \group_begin:
  \leavevmode
  \tl_if_exist:cTF{file_\file_mdfive_hash:n{#3}}{
    \tl_set_eq:Nc\video{file_\file_mdfive_hash:n{#3}}
  }{
    \IfFileExists{#3}{}{\GenericError{}{File~`#3'~not~found}{}{}}
    \pbs_pdfobj:nnn{}{fstream}{{}{#3}}
    \pbs_pdfobj:nnn{}{dict}{
      /Type/Filespec/F~(#3)/UF~(#3)
      /EF~<</F~\pbs_pdflastobj:>>
    }
    \tl_set:Nx\video{\pbs_pdflastobj:}
    \tl_gset_eq:cN{file_\file_mdfive_hash:n{#3}}\video
  }
  \pbs_pdfobj:nnn{}{dict}{
    /Type/RichMediaInstance/Subtype/Video
    /Asset~\video
    /Params~<</FlashVars (
      source=#3&
      skin=SkinOverAllNoFullNoCaption.swf&
      skinAutoHide=true&
      skinBackgroundColor=0x5F5F5F&
      skinBackgroundAlpha=0
    )>>
  }
  \pbs_pdfobj:nnn{}{dict}{
    /Type/RichMediaConfiguration/Subtype/Video
    /Instances~[\pbs_pdflastobj:]
  }
  \pbs_pdfobj:nnn{}{dict}{
    /Type/RichMediaContent
    /Assets~<<
      /Names~[(#3)~\video]
    >>
    /Configurations~[\pbs_pdflastobj:]
  }
  \tl_set:Nx\rmcontent{\pbs_pdflastobj:}
  \pbs_pdfobj:nnn{}{dict}{
    /Activation~<<
      /Condition/\IfBooleanTF{#1}{PV}{XA}
      /Presentation~<</Style/Embedded>>
    >>
    /Deactivation~<</Condition/PI>>
  }
  \hbox_set:Nn\l_tmpa_box{#2}
  \tl_set:Nx\l_box_wd_tl{\dim_use:N\box_wd:N\l_tmpa_box}
  \tl_set:Nx\l_box_ht_tl{\dim_use:N\box_ht:N\l_tmpa_box}
  \tl_set:Nx\l_box_dp_tl{\dim_use:N\box_dp:N\l_tmpa_box}
  \pbs_pdfxform:nnnnn{1}{1}{}{}{\l_tmpa_box}
  \pbs_pdfannot:nnnn{\l_box_wd_tl}{\l_box_ht_tl}{\l_box_dp_tl}{
    /Subtype/RichMedia
    /BS~<</W~0/S/S>>
    /Contents~(embedded~video~file:#3)
    /NM~(rma:#3)
    /AP~<</N~\pbs_pdflastxform:>>
    /RichMediaSettings~\pbs_pdflastobj:
    /RichMediaContent~\rmcontent
  }
  \phantom{#2}
  \group_end:
}
\ExplSyntaxOff